\documentstyle[12pt]{article}
\def\beqa{\begin{eqnarray}}
\def\eeqa{\end{eqnarray}}
\def\beq{\begin{equation}}
\def\eeq{\end{equation}}
\begin{document}
\def\bib#1{[{\ref{#1}}]}
\def\at{\tilde{a}}
\setcounter{page}{2}

\def\CcC{{\hbox{\tenrm C\kern-.45em{\vrule height.67em width0.08em depth-.04em
\hskip.45em }}}}
\def\RrR{{\hbox{\tenrm I\kern-.17em{R}}}}
\def\HhH{{\hbox{\tenrm {I\kern-.18em{H}}\kern-.18em{I}}}}
\def\DdD{{\hbox{\tenrm {I\kern-.18em{D}}\kern-.36em {\vrule height.62em 
width0.08em depth-.04em\hskip.36em}}}}
\def\ZzZ{{\hbox{\tenrm Z\kern-.31em{Z}}}}
\def\IiI{{\hbox{\tenrm I\kern-.19em{I}}}}
\def\NnN{{\hbox{\tenrm {I\kern-.18em{N}}\kern-.18em{I}}}}
\def\QqQ{{\hbox{\tenrm {{Q\kern-.54em{\vrule height.61em width0.05em 
depth-.04em}\hskip.54em}\kern-.34em{\vrule height.59em width0.05em depth-.04em}}
\hskip.34em}}}
\def\OoO{{\hbox{\tenrm {{O\kern-.54em{\vrule height.61em width0.05em 
depth-.04em}\hskip.54em}\kern-.34em{\vrule height.59em width0.05em depth-.04em}}
\hskip.34em}}}

\def\Csi{\Xi}
\def\id{{\rm id}}
\def\uno{{\bf 1}}
\def\uq2{U_q({\uit su}(2))}

\def\S{\Sigma}
\def\sh{{\rm sh}}
\def\ch{{\rm ch}}
\def\ad{A^{\dagger}}
\def\ac{a^{\dagger}}
\def\bd{B^{\dagger}}

\def\som#1{\sum\limits_{#1}}
\def\somma#1#2#3{\sum\limits_{#1=#2}^{#3}}
\def\integrale{\displaystyle\int\limits_{-\infty}^{+\infty}}
\def\intlims#1#2{\int\limits_#1^#2}
\def\tendepiu#1{\matrix{\phantom{x}\cr \longrightarrow\cr{#1\rightarrow +\infty
               \atop\phantom{x}}\cr}}
\def\tendezero#1{\matrix{\phantom{x}\cr \longrightarrow\cr{#1\rightarrow 0
               \atop\phantom{x}}\cr}}
\def\num{\strut\displaystyle}
\def\den{\over\displaystyle}
\def\fraz#1#2{{\strut\displaystyle #1\over\displaystyle #2}}
\def\bra#1{\langle~#1~|}
\def\ket#1{|~#1~\rangle}
\def\ave#1{\left\langle\left\langle~#1~\right\rangle\right\rangle}
\def\op#1{\hat#1}
\def\scalare#1#2{\langle~#1~|~#2~\rangle}
\def\esp#1{e^{\displaystyle#1}}
\def\qsp#1{q^{\displaystyle#1}}
\def\derpar#1#2{\fraz{\partial#1}{\partial#2}}
\def\part#1{\fraz{\partial}{\partial#1}}
\def\tg{{\rm tg}}

\def\ii#1{\item{$\phantom{1}#1~$}}
\def\ij{\item{$\phantom{11.}~$}}
\def\jj#1{\item{$#1.~$}}

\def\tens{\otimes}
\def\per{{\times}}
\def\co{\Delta}

\def\su2q{SU(2)_q}
\def\h1q{H(1)_q}
\def\edq{E(2)_q}
\def\etq{E(3)_q}
\def\so{SO(4)}
\def\soq{SO(4)_q}
\def\al#1{{\alpha}_{#1}}

\def\jt{J_3}
\def\jpm{J_{\pm}}
\def\jp{J_{+}}
\def\jm{J_{-}}
\def\ju{J_{1}}
\def\jd{J_{2}}
\def\j#1{J_{#1}}

\def\k#1{K_{#1}}

\def\nt{N_3}
\def\npm{N_{\pm}}
\def\np{N_{+}}
\def\nm{N_{-}}
\def\nu{N_{1}}
\def\nd{N_{2}}
\def\n#1{N_{#1}}

\def\pt{P_3}
\def\ppm{P_{\pm}}
\def\pp{P_{+}}
\def\pmen{P_{-}}
\def\pu{P_{1}}
\def\pd{P_{2}}
\def\p#1{P_{#1}}
\def\pol#1#2#3{{\cal P}_#1(#2,#3)}

\def\ee{\varepsilon}
\def\em{\ee^{-1}}

\def\e#1{{\bf e}_{#1}}

\def\epz#1{\esp{zJ_3^{#1}/2}}
\def\emz#1{\esp{-zJ_3^{#1}/2}}
\def\epw{\esp{wP_3/2}}
\def\emw{\esp{-wP_3/2}}
\def\epv{\esp{vM/2}}
\def\emv{\esp{-vM/2}}

\def\bin#1#2{\left[{\strut\displaystyle #1\atop\displaystyle #2}\right]_q}
\def\mapdown#1{\Big\| \rlap{$\vcenter{\hbox{$\scriptstyle#1$}}$}}
\def\mapcup{\bigcup}
\def\iinn{\rlap{{$\bigcup$}{$\!\!\! |$}}} 
\def\mapin{\iinn}

\begin{titlepage}
\title{Quantum Groups and Von Neumann Theorem}

\author{Alfredo Iorio and Giuseppe Vitiello 
\\{\it Dipartimento di Fisica - Universit\`a di Salerno and INFN--Napoli}
\\{\it I84100 Salerno, Italy}
\\{E--mail: iorio@sa.infn.it vitiello@sa.infn.it}
\\
\\{\bf Paper published in Mod. Phys. Lett. B {\bf 8} (1994) 269}}
              \date{\empty}
              \maketitle

\begin{abstract}
\noindent              
We discuss the $q$ deformation of Weyl-Heisenberg algebra in 
connection with the von Neumann theorem in Quantum Mechanics.
We show that the $q$-deformation parameter labels the 
Weyl systems in Quantum Mechanics and the unitarily inequivalent
representations of the canonical commutation relations  
in Quantum Field Theory. 
\end{abstract}

\thispagestyle{empty} \vspace{20. mm}


              \vfill
          \end{titlepage}

\section{Introduction}

\noindent
The von Neumann theorem in Quantum Mechanics (QM) 
states that for systems with 
finite number of degrees of freedom the representations 
of the canonical commutation 
relations (ccr), also called the Weyl representations, 
are each other unitarily equivalent. 
In the limit of infinite number of degrees of freedom, namely in 
Quantum Field Theory (QFT),
the space of physical states splits into 
unitarily inequivalent representations\cite{Bog}.
Thus in QFT, contrarily to QM, 
one has to single out the appropriate representation 
to properly describe the 
system under study:
different, i.e. unitarily inequivalent, 
representations describe different physical 
situations.
Typical examples are the ones related with 
spontaneous breakdown of symmetry 
where different $phases$ are described by different representations, e.g. 
the superconducting phase and the normal phase are described by states 
belonging to orthogonal (unitarily inequivalent) Hilbert spaces: 
the Bogoliubov 
transformations in superconductivity are (non-unitary) canonical
transformations
which relate such different representations\cite{Ume} 
(of course, one works at finite 
volume where the transformations are formally well 
defined and at the end of the computations
one goes to the infinite volume limit). 

\noindent
In  QFT the representations of the ccr may be generally 
labelled by some physically significant parameter, e.g. 
in spontaneously broken symmetry theories different
values of the order
parameter are associated to different representations;
in thermal
field theories temperature labels inequivalent 
representations\cite{Ume}; in the case of dissipative systems 
the label is the time 
variable which thus play the role of a parameter\cite{Cel}. 

\noindent
In this paper our 
purpose is to show that $q$-deformations of the Weyl-Heisenberg algebra 
($q$-WH)\cite{Bie} do indeed introduce a labelling of the 
Weyl representations in QM and of the inequivalent representations 
in QFT, the label being the $q$-deformation parameter. 

\noindent
The fact 
that $q$-WH algebra is in this way associated with the 
foliation of the space of the states in QFT 
is remarkable since some ligth may thus be
shed on the physical meaning of the deformation parameter
in view of the physical content of the representations;
moreover, $q$-parametrization may signal the presence of some finite
spacing in the theory since
$q$-WH algebra has been shown\cite{Cel2} to be related with 
systems characterized by some finite
(space or time) scale. 
The relation of inequivalent representations with $q$-WH
algebra is also interesting from the mathematical point of view
since the rich mathematical structure underlying $q$-algebras is thus
recognized to underly QFT, too.

\noindent
The properties of $q$-algebras will not be discussed in this paper; 
we simply remind that they are 
deformations of the enveloping algebras of Lie algebras and have
Hopf algebra structure\cite{Bie}. In particular, the $q$-WH algebra has the
properties of Hopf superalgebra\cite{Cel3}. The interest in $q$-algebras
arose in non-linear dynamical systems as well as
in statistical mechanics, in conformal theories,  
in solid state physics, etc..  
$q$-WH algebra has been studied in the framework of the entire
analytic functions theory and has been related with theta functions
and with Bloch functions\cite{Cel2}.
In ref. \cite{Ior} $q$-WH algebra has ben related with dissipative systems
in QFT and with thermal field theories. 

\noindent
In order to show that the $q$-deformation parameter labels the
inequivalent representations in QFT, we need first to study the Weyl
representations and the von
Neumann theorem in QM; this is done in the following section 2.
The realization of $q$-WH algebra in terms of finite difference
operators\cite{Cel2} and its relation with the generator of Bogoliubov
transformations are presented in section 3. 
The parametrization of the Weyl representations by
the $q$-parameter and its extension to the infinite volume limit in
QFT is finally established in section 4.

\section{The von Neumann uniqueness theorem} 

\noindent
In this section we present the general formalism to 
introduce von Neumann 
theorem in QM and we show how to parametrize
the representations of the ccr. 

\noindent
Let us consider a system of $M$ degrees of freedom. 
By setting 
${\hbar} = 1$,
the position and the momentum operators are
introduced as usual in the Schr\"odinger representations as 
${\hat x}_j~\to~{x}_j$ and 
${\hat p}_j~\to~-i~{d\over{d {{x}_{j}}}}$, 
$j=1,2,\ldots,M$, respectively, 
with commutation relations
\beq
[{\hat {x}}_j,{\hat p_{k}}]~=~i \delta_{j,k}~,
~~[{\hat {x}}_j,{\hat {x}}_k]~=~0~,~
~~[{\hat p_{j}},{\hat p_{k}}]~=~0~,
 \quad ~~~
\forall j,k 
\eeq
The operators of creation and annihilation are
\beq
c_j~\equiv~{{1\over{\sqrt{2}}}({\hat {x}}_j+i~{\hat p_{j}})},~~~~
c^{\dagger}_j~\equiv~{{1\over{\sqrt{2}}}({\hat {x}}_j-
i~{\hat p_{j}})}, 
\eeq
with commutators
\beq
[c_j,c^{\dagger}_k]~=~\delta_{j,k}~, 
~~[c_j,c_k]~=~0~,
~~[c^{\dagger}_j,c^{\dagger}_k]~=~0~,
\quad~~~ \forall j,k 
\eeq
As well known, since ${\hat x}$ and ${\hat p}$ 
are unbounded operators, it is necessary 
to introduce the $Weyl~ system$ of unitary operators
\beq
U({\alpha})~\equiv~\exp{\left (i\sum_{j=1}^M\alpha_j
{\hat p}_j \right)},~~~
V({\beta})~\equiv~\exp{\left (i\sum_{j=1}^M\beta_j{\hat {x}}_j \right)}
\eeq
with $\alpha,\beta \in {\cal R}^M, \alpha \equiv (\alpha_{1},...,\alpha_
{M}), \beta \equiv (\beta_{1},...,\beta_{M})$.
The ccr (1) (and (3)) are then represented by
\beq
U({\alpha})U({\alpha}')=U({\alpha}+{\alpha}')~,~
V({\beta})V({\beta}')=U({\beta}+{\beta}')~,~
U({\alpha})V({\beta})={\rm e}^{i\alpha\beta}V({\beta})U({\alpha})
\eeq
In terms of the {\it Weyl operators} $W(z)$,
\beq
W(z)~\equiv~W({\alpha}+ i {\beta})={\rm e}^{i\alpha\beta}V(
\sqrt{2}{\alpha})U(\sqrt{2}{\beta})
\eeq
with the complex variable 
$z \equiv \alpha + 
i \beta \in {\cal C}^{M}$, eqs. (5) lead to
\beq
W(z_1)W(z_2)=\exp{\left (-i Im (z^{*}_1~z_2) \right)} W(z_1+z_2)
\eeq

In the transition from QM to QFT, i.e. from 
finite to infinite number of
degrees of freedom, one 
must operate on the complex linear space ${\cal E}_C = 
{\cal E} + i {\cal E}$,
instead of working on ${\cal C}^M$. Here
${\cal E}$ denotes a real linear space of square-integrable 
functions $f$; we will denote by $F = f + i g$~,~ 
$f~,~g \in {\cal E}$~,~ the 
elements of ${\cal E}_C$. The scalar product $\langle F_1~,~F_2 \rangle$
in ${\cal E}_C$~ is defined through the scalar product $(f~,~g)$ 
in ${\cal E}$~:
\beq
\langle F_1~,~F_2 \rangle~\equiv~(f_1+i g_1~,~f_2+i g_2)=
(f_1~,~f_2) + 
(g_1~,~g_2) + i [(f_1~,~g_2)-(f_2~,~g_1)]
\eeq
In QFT the Weyl operators and their algebra (6) and (7) become
\beqa
W(F)~\equiv~W(f + i g)&=& {\rm e}^{i(f~,~g)}V(\sqrt{2} f)U(\sqrt{2} g)
\nonumber \\
W(F_1)W(F_2) &= &\exp{\left (-i Im \langle F_1~,~F_2 \rangle \right )}
W(F_1+F_2)
\eeqa
We stress that the use of complex linear space  ${\cal E}_C$ in QFT is 
required to 
{\sl smear out}  spatial integrations of field operators 
by means of {\sl test functions}~ $f$. 

\noindent
The von Neumann uniqueness theorem in QM
can be now stated as follows [1,8]:
{\it Each Weyl system with finite number $M$ of degrees of 
freedom is unitarily equivalent to the Schr\"odinger representation.
Each reducible Weyl system with finite number $M$ of 
degrees of freedom is the 
direct sum of 
irreducible representations; hence it is a multiple of the Schr\"odinger 
representation.}

\noindent
We observe that unitarily equivalence among the Weyl systems is lost
in the limit of infinite number of 
degrees of freedom, $M \rightarrow \infty$: thus
in QFT the Weyl systems provide unitarily 
inequivalent representations of the ccr.

\noindent
Let us consider now the problem of labelling each Weyl system in a way
to preserve the Weyl algebra. To this aim we consider
the following transformations
\beq
{\alpha}_i \mapsto \rho {\alpha}_i,~~~
{\beta}_i \mapsto {1\over{\rho}} {\beta}_i, \quad i=1,2,~~~\rho \neq 0
\eeq
These are canonical transformations since $Im (z^{*}_1 z_2)$
is left invariant under them and the Weyl algebra (6) is therefore 
preserved:
\beq
W^{(\rho)}(z_1)W^{(\rho)}(z_2)=\exp{\left (-i Im (z*_1z_2) \right )}
W^{(\rho)}(z_1+z_2),~~~\rho \neq 0
\eeq
where $W^{(\rho)}(z)~\equiv~W(\rho \alpha + i{1\over{\rho}}\beta)$.
We thus conclude that the transformation parameter $\rho$ labels 
the Weyl systems. 

\noindent
In QFT, i.e. for 
$M \rightarrow \infty$,
we use the canonical
transformations
\beqa
g_j \to \rho~g_j~~~,f_j \to {1\over{\rho}}~f_j,~~~\rho \neq 0~
\nonumber \\
Im \langle F_1~,~F_2 \rangle \rightarrow Im \langle F_1~,~F_2 \rangle  
\eeqa
and, in such a limit,
the representations $W^{(\rho)}(F)~,~W^{(\rho')}(F)~,~ \rho \neq \rho'$, are  
each other unitarily {\sl inequivalent}[1,8,9].

\section{$q$-Weyl-Heisenberg algebra}

\noindent
In this section we establish the relation between the $q$-WH
algebra and the Bogoliubov transformations to be used in section 4.

\noindent
From now on, for simplicity we limit ourself to one 
degree of freedom ($M=1$); 
extension to many (finite) degrees of freedom is 
straightforward.
Let us consider the Weyl-Heisenberg (WH) algebra 
realized in terms of the set of operators $\{ a~,~  a~,~N~ \}$,  
with relations:
\beq
[a,a^\dagger]=\IiI~,~~[N,a]=-a~,~~[N,a^\dagger]=a^\dagger 
\eeq
The $q$-{\sl deformation} of the WH ($q$-WH) algebra realized in 
terms of the set of operators $\{ a_q, {\hat a}_q, N_q \equiv N~ ;
 ~q~=~{\rm e}^{\epsilon} \in {\cal C} \}$, with  
$a_q ~\to~ a,~~{\hat a}_q ~\to~ a^\dagger$ as $q ~\to~ 1$, is\cite{Bie}
\beq
[a_q,{\hat a}_q]=q^N~,~~[N,a_q]=-a_q~,~~[N,{\hat a}_q]={\hat a}_q
\eeq
Since in our study of the $q$-deformation we want
to preserve the analytic properties of the Lie algebra structure, 
it is useful to work with the space ${\cal F}$ of the (entire) 
analytical functions. We therefore will adopt the
Fock-Bargmann representation (FBR)\cite{Per} where the operators
\beq
a^\dagger ~\mapsto~ \zeta~,~~a ~\mapsto~ {d\over{d\zeta}}~,
~~N \mapsto \zeta {d\over{d\zeta}}~,~ \qquad  
\zeta \in {\cal C}
\eeq
provide a realization of the WH algebra (13). 
The $q$-WH algebra (14)is realized by the operators\cite{Cel2}
\beq
{\hat a}_q ~\to~ \zeta~,~~ a_q ~\to~ {\cal D}_q~,~~
N \mapsto \zeta {d\over{d\zeta}}
\eeq
with the finite difference operator ${\cal D}_q$ (also 
called the $q$-derivative operator)
defined by
\beq
{\cal D}_q~f(\zeta)~=~{{f(q~\zeta)-f(\zeta)} \over {(q-1)~\zeta}}~,~~ 
q \neq 1~,~~ f \in {\cal F}
\eeq
Clearly, eq. (17) gives the
usual derivative in the limit of the deformation 
parameter $\epsilon$ going to zero (or $q~ \to~ 1$), 
and in the same limit the FBR (15) is obtained from (16).
One readily obtains
\beq
[a_q,{\hat a}_q]~f(\zeta)~=~q^N~f(\zeta)~=~f(q~\zeta)~,~~
 f \in {\cal F}
\eeq
The $q$-deformation of the WH algebra is therefore 
strictly related with the 
finite difference operator ${\cal D}_q$ ($q \neq 1$). 
This suggests to us that the $q$-deformation 
of the operator algebra 
should arise in the presence 
of lattice or discrete structures\cite{Cel2}.

\noindent
In terms of the operators $\tilde c$ and 
${\tilde c}^{\dagger}$ defined by
\beq
{\tilde c}~\equiv~{{1\over{\sqrt{2}}}({\hat {\zeta}}
+i~{\hat p_{\zeta}})},~~~~
{\tilde c}^{\dagger}~\equiv~{{1\over{\sqrt{2}}}({\hat {\zeta}} -
i~{\hat p_{\zeta}})}
\eeq
with ~${\hat p_{\zeta}} = -i {d \over{d{\zeta}}}$ and 
usual ccr, we have
\beq
{\zeta}~{d \over{d{\zeta}}}f(\zeta)~=~{\left [{1\over{2}}~
({\tilde c}^2 - {{\tilde c}^{\dagger 2}})~-~{1\over{2}} \right ]}f(\zeta)~,
~~~f \in {\cal F}
\eeq
and, in ${\cal F}$,
\beq
[a_q,{\hat a}_q]~=~q^N~=~{1\over \sqrt{q}}~
\exp{\left ({\epsilon\over{2}}~
({\tilde c}^2 - {{\tilde c}^{\dagger 2}}) \right)}~\equiv~{1\over \sqrt{q}}
~S(\epsilon)~,~~~q \neq 0
\eeq
Eq. (21) shows that, for $\epsilon~ \in {\cal R}$~, 
${\sqrt{q}}~[a_q,{\hat a}_q] \equiv S(\epsilon)$
is the generator of the 
Bogoliubov transformation
\beqa
{\tilde c} ~\to~{\tilde c}(\epsilon)~=~S^{-1}(\epsilon)~{\tilde c}
~S(\epsilon)&=&
{\tilde c}~cosh~\epsilon~ -{\tilde c}^\dagger~sinh~\epsilon \nonumber \\
&& \\
{\tilde c}^{\dagger}~\to~ {\tilde c}^{\dagger}(\epsilon)~=~
S^{-1}(\epsilon)~{\tilde c}^{\dagger}
~S(\epsilon)&=&
{\tilde c}^{\dagger}~cosh~\epsilon~ - {\tilde c}~sinh~\epsilon \nonumber
\nonumber
\eeqa
We close this section by observing that in ref. [5] it has been shown that 
${\sqrt{q}}~[a_q,{\hat a}_q]$ also acts as the generator of the
squeezed coherent states\cite{Yue}.

\section{$q$-WH algebra and the von Neumann theorem}

\noindent
By using the result (21) and (22) it is easy now to show that the
parameter $q$ labels the Weyl systems in QM and the inequivalent
representations in QFT.

\noindent
We consider the Weyl operator (6) (for $M~=~1$ for simplicity)
and parameterize the representations by implementing (10). We
observe that due to the definitions (4)
the transformations (10) can be equivalently 
thought as applied to 
${\hat x}$ and ${\hat p}$ instead than to $\alpha$ and $\beta$ 
(or to $f$ and $g$):
\beq
{\hat x}~ \mapsto ~{\hat {x}}(\rho) ~\equiv~\rho~{\hat x},
~~~~{\hat p}~\mapsto~{\hat {p}}(\rho)~\equiv~
{1\over{\rho}}~{\hat p},\quad \rho \neq 0
\eeq
The {\sl action variable} $J = \int p d{x}$ 
is invariant under the 
transformations (22); this
clarifies the physical meaning of the invariance of the area 
$Im (z^{*}_1 z_2)$ under (10). 
Of course, (22) is a canonical transformation:
$[{\hat x},{\hat p}]~=~i~~ \Rightarrow~~
[{\hat {x}}(\rho),{\hat p}(\rho)]~=~i$~.
In terms of $c$ and $c^\dagger$ in the Schr\"odinger representation
(see eq. (2)) eqs. (22) read
\beqa
{\hat x}~ \mapsto ~{\hat x}(\rho) ~\equiv~\rho~{\hat x}
&=&{1\over{\sqrt{2}}}\left(\rho~c + 
\rho~c^\dagger \right) \nonumber \\
&& \\
{\hat p}~\mapsto~{\hat { p}}(\rho)~\equiv~
{1\over{\rho}}~{\hat p}&=&
{-i \over{\sqrt{2}}}\left({1 \over{\rho}}~c - 
{1 \over{\rho}}~
c^\dagger~\right)~,~~~\rho ~ \neq ~0 \nonumber
\eeqa
We may thus introduce the operators $c(\rho)$ and $c^{\dagger}(\rho)$
as
\beqa
c ~\mapsto~ c(\rho)~\equiv~{1 \over{\sqrt{2}}} \left({\hat x}(\rho)~+~
i{\hat p}(\rho) \right)
&=& {1\over{2}}{ \left[\left(\rho
+ {1\over{\rho}}
\right)c + \left(\rho - {1 \over{\rho}}\right)c^\dagger\right]} \nonumber \\
&& \\
c^\dagger ~\mapsto~ c^\dagger (\rho)~\equiv~
{1 \over{\sqrt{2}}}\left({\hat x}(\rho)~-~i
{\hat p}(\rho)\right)&=&
{1\over{2}}{\left[\left(\rho + 
{1\over{\rho}}\right)c^\dagger  + 
\left(\rho - {1\over{\rho}}\right)c\right]} \nonumber
\eeqa
By assuming $\rho$ real and defining
\beq
u~\equiv~{1\over{2}}\left(\rho +{1\over{\rho}}\right)~,~~
v~\equiv~{1\over{2}}\left(\rho -{1\over{\rho}}\right)
\eeq
so that $u^{2}~-~v^{2}~=~1$~, 
eqs. (25) are then recognized to be nothing but Bogoliubov 
transformations. Let 
$\rho~\equiv q^{-1}~=~{\rm e}^{-\epsilon}$~,
$\epsilon \neq \infty$ and real.
Eqs. (25) are then put in the more familiar form 
\beqa
c(\epsilon)&=&
c~cosh~\epsilon~ - c^\dagger~sinh~\epsilon \\
c^\dagger(\epsilon)&=&c^\dagger~ cosh{\epsilon}  - c~sinh{\epsilon} 
\nonumber 
\eeqa
where we used the notation ${c(\epsilon)~\equiv~c(\rho)}$~.
We thus 
reach the (well known[1,8]) conclusion that the representations $W(z)$
and $W^{(q)}~ \equiv ~W^{(\rho)}(z)$ are 
connected by Bogoliubov transformations. The $\rho$-parametrization is 
called the {\sl Bogoliubov parameterization}. 

\noindent
The connection with the $q$-WH algebra (14) is immediately seen by
realizing that from the holomorphy condition on $f(\zeta)$, 
~$f \in {\cal F}$, with 
$\zeta~=~x + iy$, we have ${d\over{d{\zeta}}}f(\zeta)
={d\over{d x}}f(\zeta)$, so that ${\hat p}_{\zeta} = {\hat p}$ 
and eqs.(22) give
\beq
~S^{-1}(\epsilon)~{\tilde c}~S(\epsilon) \to c(\epsilon) \quad {\rm and}
~S^{-1}(\epsilon)~
{\tilde c}^\dagger~S(\epsilon)~ \to~c^\dagger(\epsilon)
\eeq
as $y \to 0$.

\noindent
We therefore conclude that, in the $y \to 0 $ limit,
the $q$-WH algebra commutator 
$[a_{q},{\hat a}_{q}]$
acts (up to the $\sqrt{q}$ factor) as the generator of the Bogoliubov
transformations (27) 
relating $W(z)$ with $W^{(q)}(z)$, the $q$-deformation
parameter labelling the Weyl representations. 

\noindent
We also observe that same conclusion is  
reached by working since the beginning
in the complex $\zeta$-plane, $\zeta = x + iy$, with functions 
$f(\zeta) \in {\cal F}$ ~(in the extension to the $\zeta$-plane the
Weyl operator (6) acquires the phase factor 
$e^{-{\sqrt{2}}{\alpha}y}$ (cf. eq.(6) for $M = 1$)).

\noindent
In the limit of infinite degrees of freedom 
the representations $W^{(q)}(F)~,~W^{(q')}(F)$,
~$q \neq q'$~, are  
unitarily {\sl inequivalent}[1,8,9]. 
Different values of the $q$-deformation parameter thus 
label unitarily inequivalent 
representations in QFT.

\section{Conclusions }

\noindent
We have shown that the $q$-deformation parameter of the $q$-WH 
algebra labels the Weyl systems in QM and 
the unitarily inequivalent representations of the ccr in QFT. 
We found that the generator of the Bogoliubov transformations which
relate Weyl systems or representations labelled by different values 
of $q$ 
is (up to a c-number factor) the
$q$-WH algebra commutator $[a_q,{\hat a}_q]$. 

\noindent
The von Neumann theorem 
in QM has been discussed in relation with the 
parameterization of the Weyl 
representations through Bogoliubov transformations.

\noindent
The quantum commutation
rules (1) specify the set of canonically conjugate operators
$({\hat x}_{i}, {\hat p}_{i})$ only up to the transformations (22) 
(which are infact  canonical transformations)
even if irreducibility is assumed\cite{Ara},
so that the set of canonical operators is fully specified only when 
the value of $\epsilon = \log q$ is also given, or, in other
words, when the representation $W^{(q)}(z)$ is assigned. 
This is a trivial
problem in QM where Weyl systems with different $q$ labels are 
each other unitarily
(and therefore physically) equivalent (the von Neumann theorem). 
Not so in QFT where 
representations with different $q$ labels are 
each other unitarily inequivalent and thus the assignement
of the representation for a specific value of $q$ is physically
relevant, e.g. in spontaneously broken symmetry theories where
different physical {\it phases} of the system are described by 
different (inequivalent) representations. 
The procedure of 
tuning the $q$-parameter in the $q$-WH algebra may 
be thus understood as 
"tunneling" through inequivalent representations in QFT.

\noindent
The $q$-parametrization of 
the representations of the ccr 
provides remarkable physical interpretation of the 
$q$-deformation parameter also because $q$-WH
algebra has been related with coherent states, squeezed states,
discretized (periodic) systems\cite{Cel2}, thermal field
theories and  quantum dissipation\cite{Ior}. 

\noindent
We are glad to aknowledge useful discussions with E.
Celeghini, S.De Martino, S.De Siena and M.Rasetti.

\end{document}